\documentclass[PRL,twocolumn,amsmath,superscriptaddress]{revtex4-1}
\usepackage[colorlinks,urlcolor=blue,linkcolor=blue,citecolor=blue]{hyperref}
\usepackage{overpic}
\usepackage{subfigure}
\usepackage{latexsym}
\usepackage{braket}
\usepackage{amsfonts}
\usepackage{color,xcolor}
\usepackage{bm}
\usepackage{array}
\usepackage{booktabs}
\usepackage{caption}
\usepackage{multirow}
\newcommand{\tabincell}[2]{\begin{tabular}{@{}#1@{}}#2\end{tabular}}
\makeatletter

\newcommand{\Rmnum}[1]{\expandafter\@slowromancap\romannumeral #1@}
\makeatother

\begin{document}

	\title{Inter-orbital p- and d-wave pairings between $d_{xz/yz}$ and $d_{xy}$ orbitals in Sr$_2$RuO$_4$ }

	\author{Weipeng Chen}
	\affiliation{
		National Laboratory of Solid State Microstructures And Department of Physics, Nanjing University, Nanjing 210093, China}
	\author{Jin An}
	\affiliation{
		National Laboratory of Solid State Microstructures And Department of Physics, Nanjing University, Nanjing 210093, China}
	\affiliation{
		Collaborative Innovation Center of Advanced Microstructures, Nanjing University, Nanjing 210093, China}
	
	\date{\today}
	
	\begin{abstract}
	We study the pairing symmetry of Sr$_2$RuO$_4$ through the group-theoretical approach. We emphasize the role of pairing interaction between the quasi-one-dimensional(Q1D) $d_{xz/yz}$ and quasi-two-dimensional(Q2D) $d_{xy}$ orbitals. It is found that two degenerate inter-orbital time-reversal-invariant(TRI) p-wave pairings, one is spin-singlet and the other spin-triplet with out-of-plane $\bm{d}$-vector, could be the most promising candidates.  Several important physical quantities are presented, including the near-nodal gap structure, the unchanged out-of-plane Knight shift, and no split transition under strain, which are consistent with the experiments. In addition, these p-wave pairings shed light on resolving the contradiction between the time-reversal breaking and reduced in-plane Knight shift measurements. As the system reaches the Van Hove singularity under applied strain, the pairing symmetry would become a d-wave pairing mainly consisting of inter-orbital components, which could be responsible for the strained 3$K$ phase.
	\end{abstract}
	\maketitle
	\emph{Introduction.}-----Soon after the discovery of superconductivity($T_c=1.5 K$) in Sr$_2$RuO$_4$, plenty of experiments have been made to catch its novel physics  \cite{Ishida1998,Luke1998,Ishida2001,Laube2000,Suzuki2002,Nelson2004,Murakawa2004,Kidwingira2006,Xia2006,Jang2011}. These experiments suggest the pairing order parameter(OP) be odd-parity, spin-triplet, and time-reversal-breaking(TRB), which leads to the widely accepted $p_x + ip_y$ pairing state. However, this chiral order can't resolve some apparently contradictory results \cite{Mackenzie2003,Maeno2011,Kallin2012} such as: the nodal behaviors \cite{Graf2000,Nishizaki2000,Bonalde2000,Ishida2000,Firmo2013}; no split transition under applied symmetry-breaking fields \cite{Sigrist1987,Tsuchiizu2015,Mao2000,Yaguchi2002}, and the missing chiral edge current \cite{Kirtley2007,Curran2011,Curran2014}. It could also explain neither the Pauli limiting behavior nor the first-order transition \cite{Yonezawa2013,Kuhn2017}. Recent studies on pressure effect provide a new insight \cite{Kittaka2010,Hicks2014,Burganov2016,Steppke2017,Barber2018}. T$_c$ increases slowly with small in-plane unaxial strain, indicating the absence of split transition, and then it reaches a sharp peak ($3.4K$) at compression by $\simeq$0.6\%, accompanied by a strong enhancement of the z-axis upper critical fields $H^{\parallel}_{c2}$. A promising proposal for this is that the OP of the 3K phase is even-parity, indicating an  odd- to even-parity transition at an intermediate strain \cite{Steppke2017}. The most interesting results come from the newest NMR measurements on superconducting Sr$_2$RuO$_4$, which obtained significant drops in the in-plane Knight shift for both unstrained and strained cases, ruling out the pairings with out-of-plane $\bm{d}$-vector \cite{Pustogow2019}.

	The puzzles above have triggered huge amount of theoretical works on the pairing mechanism \cite{Hasegawa2000,Raghu2010,Scaffidi2014,Huang2016,Zhang2017,Komendov2017,Huang2018,Zhang2018,Wang2019}, but the multi-orbital nature of the material makes this issue to be rather confused and unclear. There are three bands derived from the $t_{2g}$ orbitals of Ru: $\alpha$, $\beta$ mainly from the Q1D $d_{xz/yz}$ and $\gamma$ mainly from the Q2D $d_{xy}$. Most previous studies were focused on an active Q2D orbital  \cite{Agterberg1997,Nomura2000,Wang2019}, or the Q1D orbitals \cite{Raghu2010,Chung2012,Huang2016}. We notice that pairings between the two kinds of orbitals have rarely been concerned, which could be owing to the general thought that the atomic spin-orbit couping(SOC) among the orbitals is too weak \cite{Yanase2003,Agterberg1997,Huang2016,Zhang2018}. Recent theoretical and experimental results, however, support a much stronger SOC \cite{Veenstra2014,Zhang2016M,Tamai2019}, making this kind of pairings feasible. The possibility of such pairings in Sr$_2$RuO$_4$ has been discussed very recently \cite{Puetter2012,Gingras2018,Huang2019,Kaba2019}, but detailed investigations on them are still lacking.	
		
	In this letter, we highlight the role of the pairing interaction between the Q1D and Q2D orbitals and then make an analysis on the pairing symmetries in unstrained and strained Sr$_2$RuO$_4$. We find two degenerate inter-orbital TRI p-wave pairing states, one is spin-singlet and the other spin-triplet with out-of-plane $\bm{d}$-vector. Their physical quantities are consistent with several essential experimental facts: the vertical line nodes (near nodes) in the gap, the unchanged out-of-plane and reduced in-plane Knight shift, the Pauli limiting behavior, and the absence of split transition under uniaxial strain. These p-wave pairing states also shed light on explaining the observed time-reversal breaking. Furthermore, it is found that the strained Sr$_2$RuO$_4$ would undergo an odd-to-even OP transition as the system approaches the Van Hove singularity(VHS), suggesting that an inter-orbital-dominant d-wave pairing state could be the possible candidate of the strained 3K phase.
	
	\begin{figure}
		\vspace{0.cm}
		\includegraphics[width=8.5cm]{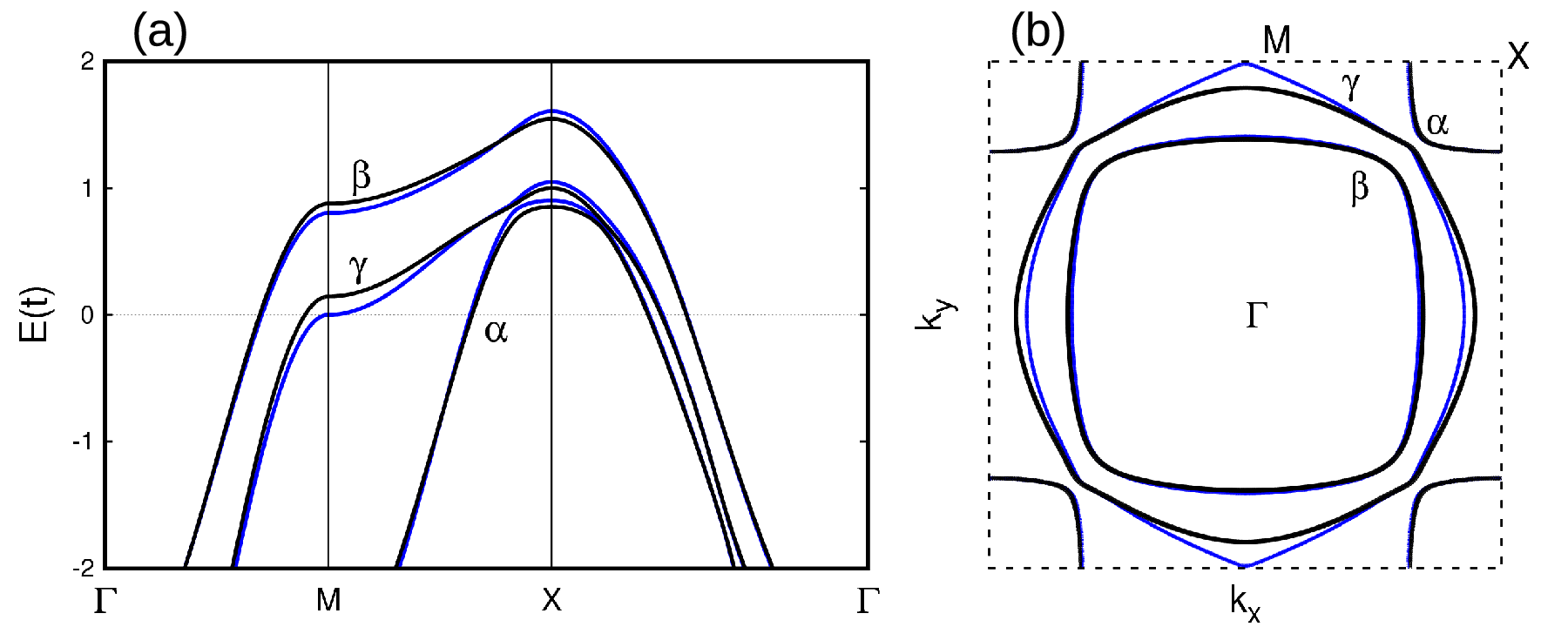}		
		\vspace{-0.2cm}
		\caption{\label{fig01}
			(a) The fitted band structure. (b)2D FS of the normal Sr$_2$RuO$_4$. The black and blue lines correspond  $\epsilon_{xx}$=0 and -0.75 \%, respectively. The SOC strength is set to be $\eta= 0.2 t$.
		}
	\end{figure}	
	\emph{Model and Method.}-----
	Based on the three orbitals, we construct a 2D tight-binding model to fit the Fermi surface(FS) revealed by the high-resolution photoemission \cite{Tamai2019,SM}. The fitted energy band and FS of Sr$_2$RuO$_4$ are exhibited in Fig. \ref{fig01}.	
	At zero strain, the 2D symmetry group of Sr$_2$RuO$_4$ could be $C_{4h}$ or $C_{4v}$. Here we adopt the latter as in Ref. \cite{Zhang2014}. We consider the $\bm{k}$-dependent pairings but ignore the on-site ones because of the deep gap minima observed in experiments \cite{Rastovski2013,Hassinger2017}. For simplicity, only the nearest-neighbor(NN) pairings are involved in our paring symmetry analysis. Then the most favorable pairing state and its $T_c$ could be determined by solving the linearized gap equation	
	\begin{eqnarray}
	\Delta_{\sigma_1 \sigma_2}(\bm{k})&=& \frac{T_c}{N}\sum\limits _{p\bm{k}^\prime \sigma_3\sigma_4} V_{\sigma_1\sigma_2\sigma_3\sigma_4}(\bm{k},\bm{k}^\prime) \nonumber\\
	&& \times[G(k^\prime)\Delta(\bm{k}^\prime)G^{\tau}(-k^\prime)]_{\sigma_3\sigma_4}
	,\end{eqnarray}
	where $G(k)=[i\omega_p-H_0(\bm{k})]^{-1}$ is the normal states' Matsubara Green functions, $\Delta(\bm{k})$ the $6\times 6$ gap matrix, $\sigma$ the index combining the three orbitals and two spins. The pairing interaction $V(\bm{k},\bm{k}^\prime)$ could be expanded by the basis gap functions(BGFs) as follows	
	\begin{eqnarray}
	V_{\sigma_1\sigma_2\sigma_3\sigma_4}&&(\bm{k},\bm{k}^\prime) = \nonumber\\
	&&\sum_{\Gamma a} V_{\Gamma,a} [D_{\Gamma,a}(\bm{k})]_{\sigma_1 \sigma_2} [D^*_{\Gamma,a}(\bm{k}')]_{\sigma_3\sigma_4}
	,\end{eqnarray}
	where $D_{\Gamma,a}(\bm{k})$ represents the $a$-th BGF for the irreducible representation(IR) $\Gamma$, with $V_{\Gamma,a}$ being the corresponding coupling strength. The gap matrix $\Delta(\bm{k})$ could thus be written as a linear superposition of the BGFs and the coefficients are determined accordingly \cite{SM}.
	
	According to the crystalline symmetry \cite{Huang2016}, $V_{\Gamma,a}$ is assumed to be independent of different IRs and only orbital dependent. It can be expressed by
	\begin{eqnarray}\label{eqn6}
	\begin{bmatrix}
	V_{xz,xz}& V_{xz,yz}&V_{xz,xy} \\
	V_{yz,xz}& V_{yz,yz}&V_{yz,xy} \\
	V_{xy,xz}& V_{xy,yz}&V_{xy,xy}
	\end{bmatrix}
	=\begin{bmatrix}
	V_1& V'_1&V_2 \\
	V'_1& V_1&V_2 \\
	V_2& V_2&V_3
	\end{bmatrix}
	,\end{eqnarray}
	where $V_1(V'_1)$ represents the intra(inter)-orbital pairing strength in the Q1D orbitals, $V_2$ between $d_{xz/yz}$ and $d_{xy}$, and $V_3$ that in $d_{xy}$, respectively. Because pairings within these two Q1D orbitals can never be dominant in our model, we simply set $V'_1=V_1$ here, similar as employed in the work of Ref. \cite{Fukaya2018}. For strained Sr$_2$RuO$_4$, $V_i$ should become anisotropic. While since the applied strains are small(denoted as $|{\epsilon_{V}}|\textless 1\%$) \cite{Hicks2014,Steppke2017,Pustogow2019}, the anisotropy is neglected so that Eq. (\ref{eqn6}) is assumed to be still valid in this situation.

	\begin{figure}
	 	\vspace{0.cm}
	 	\includegraphics[width=8.5cm]{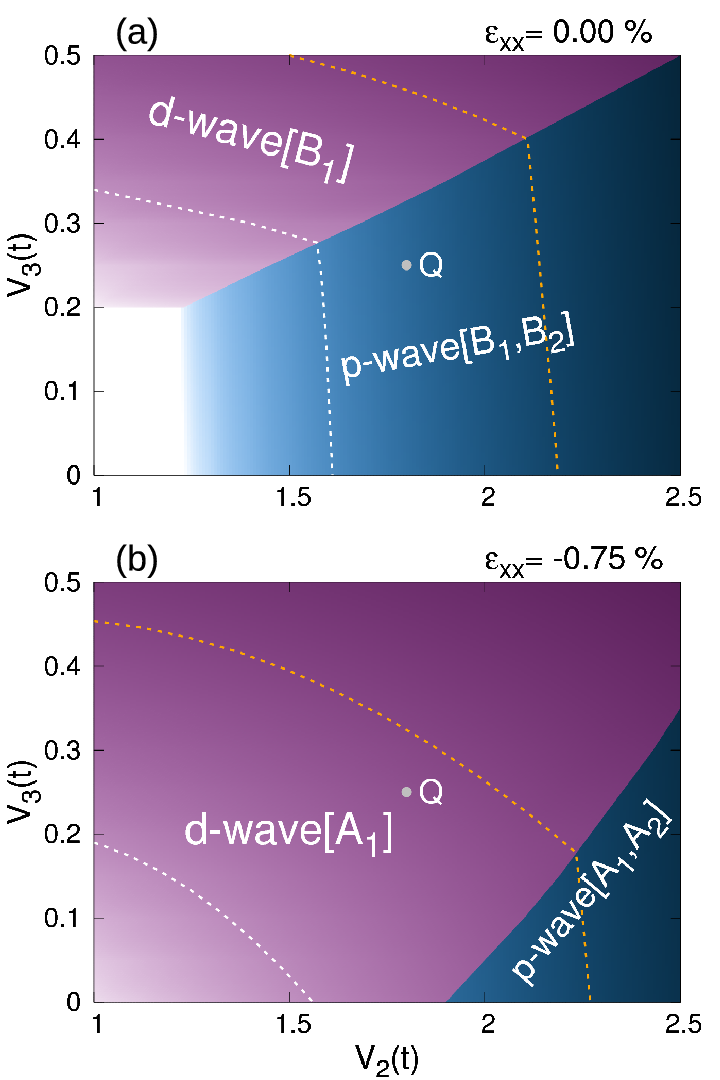}
	 	\vspace{-0.cm}
	 	\caption{\label{fig02}
	 		Pairing phase diagrams with $V_1=0.5t$. (a)Unstrained Sr$_2$RuO$_4$. The upper regime denotes the $B_1$ d-wave pairing, while the lower one denotes two degenerate p-wave pairings belonging to $B_1$ and $B_2$ of $C_{4v}$. (b)Strained Sr$_2$RuO$_4$ with $\epsilon_{xx}=\epsilon_{V}=-0.75 \%$. Due to the reduced symmetry, the d-wave pairing belongs to $A_1$, and the two p-wave pairings belong to $A_1$ and $A_2$ of $C_{2v}$. The white and orange dashed lines denote the regimes with $T_c\simeq0.5K$ and $5K$, respectively, assuming $t=0.1eV$.
	 	}
	\end{figure}	
	\emph{Phase diagram.}-----Since the inter-orbital pairings between Q1D and Q2D orbitals are expected to play an important role, we assume $V_2$ to be the dominant coupling constant, much larger than the others. Our calculations demonstrate that the phase diagram is nearly unchanged as $V_1$ varies($<0.8 V_2$), while it is very sensitive to the variation of $V_3$. Therefore, in the following we are only focused on the $V_2-V_3$ parameter space with fixed $V_1$. Fig. \ref{fig02}(a) shows the phase diagram of unstrained Sr$_2$RuO$_4$. There are two kinds of pairing states with distinct parities. One is a mixed TRI even-parity pairing state belonging to IR $B_1$ of $C_{4v}$, which is mainly composed of an inter-orbital d-wave pairing with an in-plane $\bm{d}$-vector $d_{x}^{(xz,xy)}-d_{y}^{(yz,xy)}$, and an intra-orbital d-wave pairing. Although the latter is always a secondary component \cite{SM}, it plays an essential role in stabilizing the phase. The other kind is the two degenerate TRI p-wave pairing states belonging to IR $B_1$ and $B_2$. The leading component of the $B_1$($B_2$) p-wave order is an inter-orbital singlet(triplet) BGF, denoted by $\psi^{(xz,xy)}(i \sin{k_y})$ - $\psi^{(yz,xy)}(i \sin{k_x})$ [$d_z^{(xz,xy)}( \sin{k_y}) $ - $d_z^{(yz,xy)}( \sin{k_x})$]. Their degeneracy originates from a pseudo-spin-rotation symmetry of our Hamiltonian \cite{SM}. Fig. \ref{fig02}(b) is the strained phase diagram at $\epsilon_{xx}=\epsilon_{V}$. Just like the unstrained case, only the d- and p-wave pairing states appear in the diagram. Under this unaxial strain, the point-group symmetry is reduced as $C_{4v} \rightarrow C_{2v}$ and the corresponding IRs are transformed as $\{A_1, B_1\} \rightarrow A_1$, $\{A_2, B_2\} \rightarrow A_2$ and $E\rightarrow \{B_1,B_2\}$ \cite{Ramires2019}. Although the $C_4$ symmetry breaking causes additional admixtures among the BGFs, the leading components of the two orders in Fig. \ref{fig02}(a) are still dominant in Fig. \ref{fig02}(b) \cite{SM}. The d-wave pairing now belongs to IR $A_1$ while the two degenerate p-wave pairings belong to $A_1$ and $A_2$, respectively. For the p-wave case, our following related discussions will based on $B_1$ of $C_{4v}$(unstrained) and $A_1$ of $C_{2v}$(strained).
	
	\begin{figure}
	 	\vspace{0.cm}
	 	\includegraphics[width=8cm]{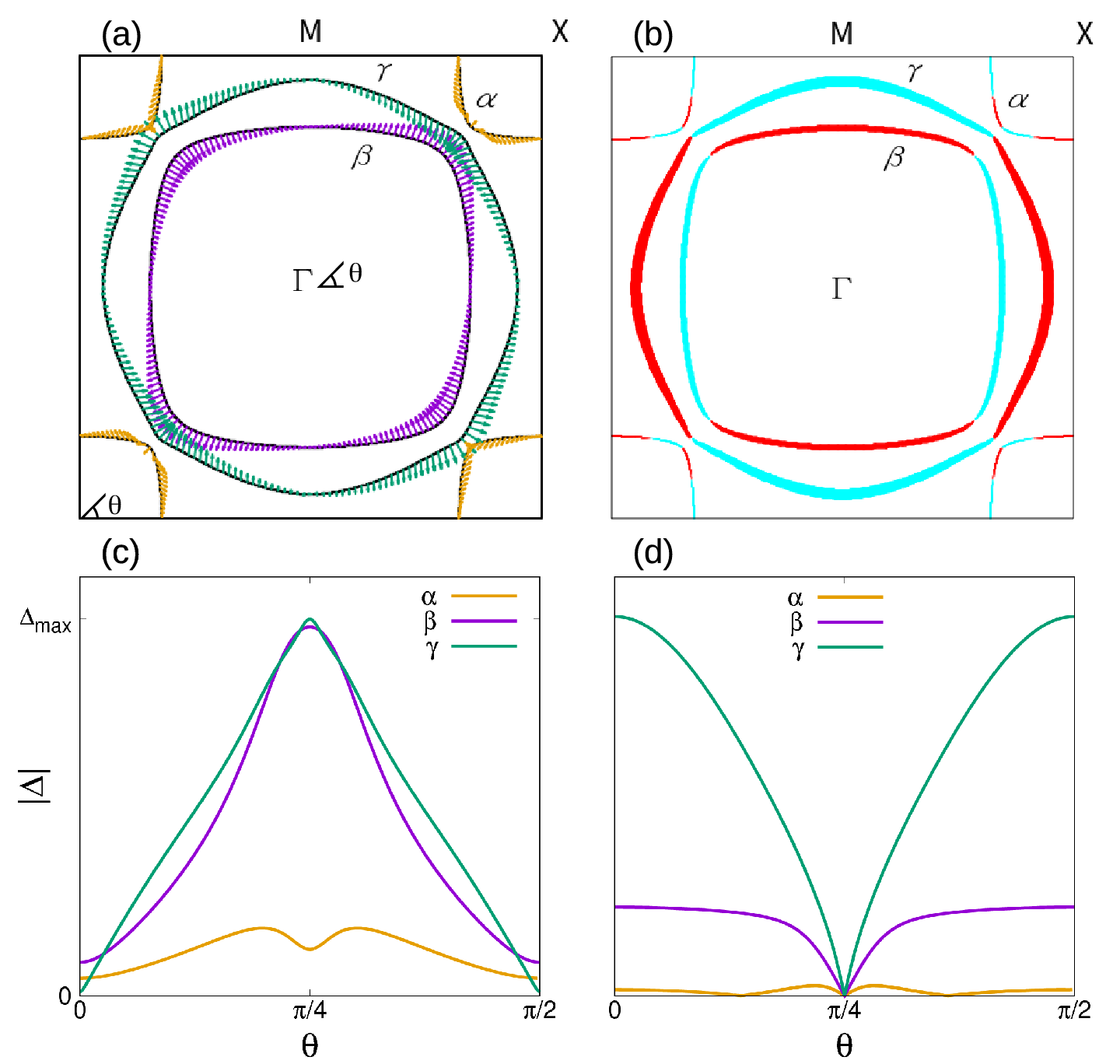}
	 	\vspace{-0.cm}
	 	\caption{\label{fig03}
			Projected order parameters on the FS in the unstrained Sr$_2$RuO$_4$, with  $\eta=0.2t$. (a) Projected $\bm{d}$-vectors of the p-wave pairing. (b) Projected $\psi$ of the d-wave pairing, where the red(cyan) line represents the gap sign $-$($+$) and the line width denotes the gap size. Their detailed gap-size dispersion on the FS are displayed in (c)(d), where $\theta$ is the azimuth angle around the FS pocket. The p-wave gaps have near nodes on all these three FS pockets at $\theta=0$, while the d-wave cases have real nodes at $\theta=\pi/4$.
	 	}
	\end{figure}	
	\emph{Unstrained Sr$_2$RuO$_4$.}-----Now we make gap projections of the two OPs without strain, and analyze their gap structures on the FS. It is found that, after the projection the p- and d-wave superconductivity remain TRI and emerge in the parallel and `anti-parallel' pseudo-spin pairing channels, respectively \cite{SM}. This allows us to describe the projected gap functions in a real $\bm{d}$-vector or $\psi$ form. The projected or pseudo-$\bm{d}$-vectors on the three bands of the p-wave pairing are in-plane as depicted by Fig. \ref{fig03}(a). Gaps open mainly on $\beta$ and $\gamma$ and have modulated helical p-wave forms, with their maxima(minima) living in the $\langle110\rangle$ ($\langle 100 \rangle$) directions. As shown in Fig. \ref{fig03}(c), all the three bands hold near nodes and the deepest ones give gaps about 1/60 of the maximum value.
	Same as we did above, projected gaps of the d-wave pairing are shown in Fig. \ref{fig03}(b). Gap $\psi^\gamma$ has the biggest magnitude and $\psi^\beta$ is much smaller, both of which take the $d_{x^2-y^2}$ form, while $\psi^\alpha$ is rather tiny. All of them hold nodes along $\langle110\rangle$, as depicted by Fig. \ref{fig03}(d).
	
	The anisotropic gaps revealed by the experiments support gap minima sitting on the $\langle100\rangle$ sections \cite{Deguchi2004}, indicating that the p-wave rather than the d-wave pairing could be a better candidate OP. To further confirm this one has to do more efforts on the physical properties for these two OPs. For the calculation details, one can refer to Ref. \cite{Wang2019}. The temperature dependent gap sizes are obtained from the self-consistent gap equation \cite{Sigrist1991,SM}.
	
\begin{figure}
		\vspace{0.2cm}
		\includegraphics[width=8.5cm]{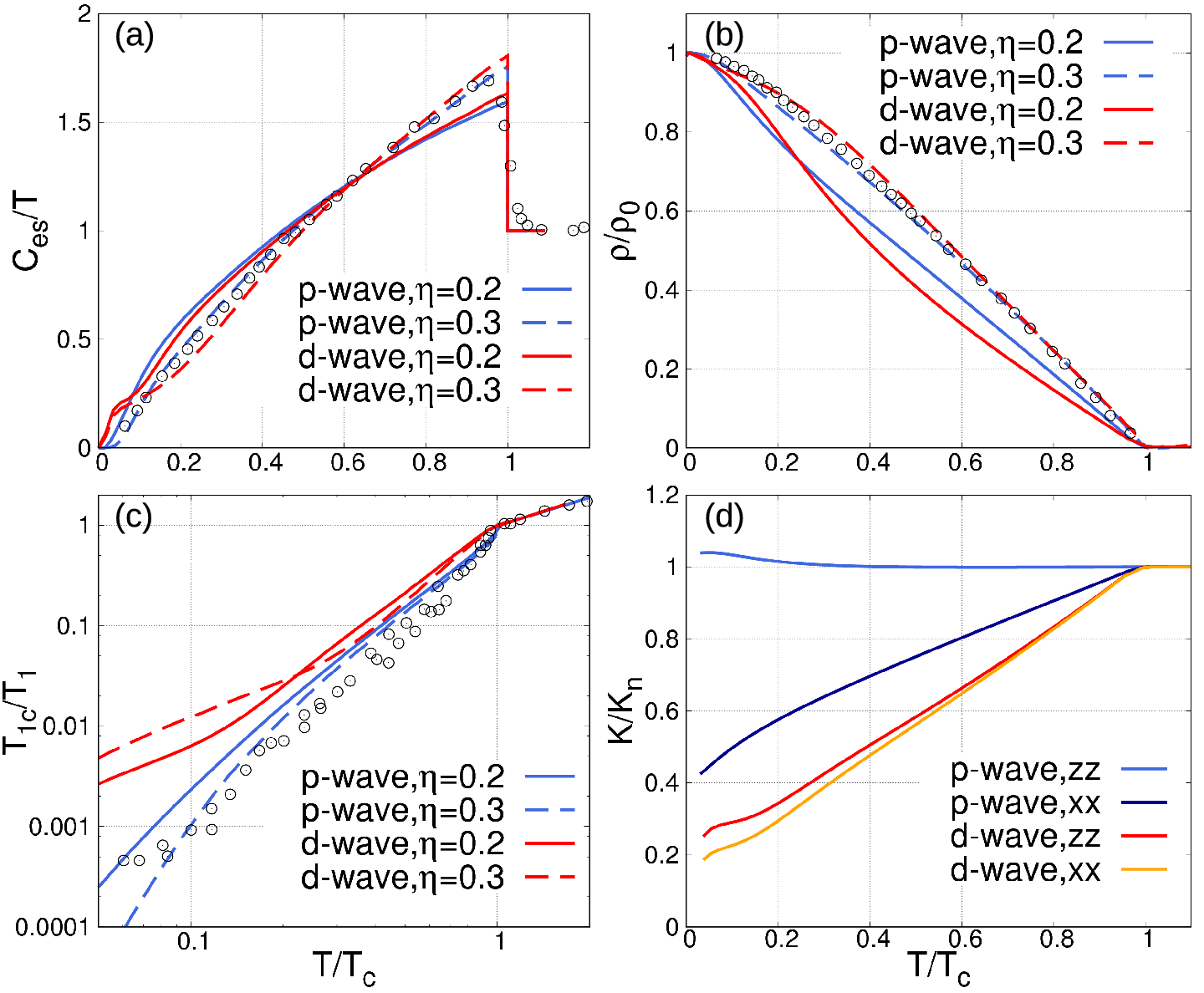}
		\vspace{-0.cm}
		\caption{\label{fig04}
			Physical properties calculated for the two OPs in Fig. \ref{fig02}(a). (a) The electronic specific heat divided by temperature $C_{es}/T$. (b) Superfluid density $\rho/\rho_0$, (c) Spin-lattice relaxation rate $T_{1c}/T_1$. The blue(red) lines represent the p-wave(d-wave) case, in which the solid(dashed) one denotes the SOC strength $\eta=0.2(0.3)t$. The circles are experimental data extracted from Ref. \cite{Nishizaki2000} for (a), Ref. \cite{Bonalde2000} for (b) and Ref. \cite{Ishida2000} for (c), respectively. (d) Normalized Knight shift $K/K_n$. The blue(red) and dark-blue(orange) lines denote $K_{zz}$ and $K_{xx}$ of the p-wave(d-wave) pairing state, respectively.
		}
	\end{figure}

	Firstly, the specific heat divided by temperature $C_{es}/T$ are shown in Fig. \ref{fig04}(a).  Both of the p- and d-wave pairings are nearly T-linear. Their specific heat jumps $\Delta C_{es}/T_c$ are 0.60 and 0.62, respectively, somewhat smaller than the experimental value 0.73 \cite{Nishizaki2000,Nomura2002,Deguchi2004}.
	At lower temperature($<0.1 T_c$), however, the d-wave pairing shows a crossover behavior owing to the tiny gap on $\alpha$ \cite{Agterberg1997}, which is incompatible with the experiments mentioned above. 	
	The superfluid density $\rho /\rho_0$ are presented in Fig. \ref{fig04}(b). To compare our results with the experiment \cite{Bonalde2000}, where the example has a $T_c= 1.39K$, an estimated elastic scattering rate $\xi=0.1T_c$ is taken into account \cite{Wang2019}. These two OPs both display a quadratic behavior just as observed by the experiment below the temperature $0.1 T_c$. Near $T_c$, the d-wave pairing has a lower slope than the p-wave one, leading to an obvious downward concave character at about $T=0.5 T_c$.	
	Fig. \ref{fig04}(c) displays the spin-lattice relaxation rate.  Comparing with the experiment in Ref. \cite{Ishida2000}, where $1/T_1 \sim T^3$ indicates a nodal structure, $T_{1c}/T_1$ of our p-wave pairing shows a slight left shift, while that of the d-wave pairing yields a large deviation at the temperature $T<0.2 T_c$. This is because the smallness of gap $\psi^\alpha$ contributes a Korringa law $1/T_1\sim T$, just like the normal state.	
	It can be seen that the theoretical results of the p-wave pairing fits the experimental data better than the d-wave case, but still not good enough. We repeat these calculations using a larger SOC $\eta=0.3t$, with parameters modified to keep the nearly unchanged FS. Notice that the fit between our results and experiments for both the p- and d-wave pairings(the dashed blue and red lines in Fig. \ref{fig04}) are improved. However, the d-wave $T_{1c}/T_1$ keeps far away from the experimental data. It can be seen that to fit these experimental data best an appropriate strong $\eta$ is required, which is coincident with our starting point.

	Fig. \ref{fig04}(d) presents the normalized Knight shift $K/K_n$, with $K_n$ being the normal value. In the d-wave case, $K_{zz}$ and $K_{xx}$ reduce as the temperature go through $T_c$. This is similar to the behavior of a one-band d-wave model and qualitatively consistent with some theoretical results \cite{Yu2018,Roising2019}.
	In the p-wave case, $K_{zz}$ shows nearly no change below $T_c$ but slightly increases near $T=0$, while $K_{xx}$ has a reduction about 58 \% there. These characters are robust with respect to a small SOC variation and in agreement with the out-of-plane Knight shift experiments \cite{Ishida2001,Murakawa2004,Ishida2015} and very recent in-plane NMR measurements \cite{Pustogow2019,Ishida2019}. We also apply the calculations on the $B_2$ p-wave pairing $d_z^{(xz,xy)} - d_z^{(yz,xy)}$, and obtain exactly the same results. Notice that even these two degenerate p-wave pairings have different spin angular momentum: one is spin-singlet and the other spin-triplet, their projected $\bm{d}$-vectors are exactly the same: both in-plane and have a nodal structure(One can check this through the projection in \cite{SM}). Furthermore, our results are in agreement with the calculation of a helical pairing defined on the FS in Ref. \cite{Roising2019}. This indicates that the Knight shift is strongly related to the projected $\bm{d}$-vectors. Given that a reduced Knight shift is always consistent with a Pauli limiting behavior, the projected $\bm{d}$-vectors could also be useful in describing the suppression of the in-plane upper critical field \cite{Rastovski2013,Amano2015,Ramires2016,Zhang2018}. In this sense, the projected gap functions could be treated as effective ones. It's interesting to find that the original odd-parity OP, which is singlet or triplet with $\bm{d} \parallel \hat{z}$, will result in an unchanged $K_{zz}$, whereas the original even-parity OP, mainly triplet with in-plane $\bm{d}$-vector, gives sharply reduced Knight shift in any directions. These findings challenge the general concept of the relation between the pairing symmetry and Knight shift, indicating distinct magnetic-field responses of the inter-orbital pairings.

	\emph{Strained Sr$_2$RuO$_4$.}-----Now we turn to study the pairing OP of the strained Sr$_2$RuO$_4$. Stain drives the 1D IRs of $C_{4v}$ to merge, which will induce additional mixing between the BGFs. For instance, the BGF $\psi^{(xz,xy)}-\psi^{(yz,xy)}$ will mix together with small amount of $\psi^{(xz,xy)}+\psi^{(yz,xy)}$, resulting in a strained p-wave pairing which could be roughly denoted by $\Delta^1\psi^{(xz,xy)}- \Delta^2\psi^{(yz,xy)}$ with $\Delta^1 \neq \Delta^2$ being different amplitudes of these two components. This demonstrates that there's no splitting of the components under strain for our p-wave case, compatible with the observation \cite{Hicks2014,Steppke2017,Pustogow2019}. The projected $\bm{d$}-vectors of the strained p-wave order remain in-plane and nearly nodal, therefore the physical properties are qualitatively similar to the unstrained case and not presented here.	
	The same analysis applied on the strained d-wave order gives projected gaps dominating on $\beta, \gamma$.
	For a set of fixed coupling parameters $V_i$, as the strain increases to be approaching $\epsilon_V$, a large area of the parameter space will go through an odd-to-even phase transition. Take the regime labeled by ``Q" in Fig. \ref{fig02} as an example. Our calculation gives a p-wave superconductor with $T_c\simeq 1.52 K$ there at zero strain, but it changes to the d-wave with enhanced
	 $T_c\simeq3.44 K$ at the compression by $\epsilon_{xx}=\epsilon_{V}$. This result is in agreement with the observed phenomenology in the strained $3 K$ Sr$_2$RuO$_4$ \cite{Steppke2017}.
	
	\emph{Discussions.}-----Lots of puzzles in Sr$_2$RuO$_4$ could be resolved by our inter-orbital pairing model, whereas it seems that the TRI p-wave pairing itself contradicts the TRB experiments \cite{Luke1998,Xia2006,Wang2017}. To reconcile this we assume that the two degenerate p-wave pairings could coexist in distinct domains. The time-reversal symmetry is broken when the relative phase between them is nontrivial. This symmetry-breaking mechanism have been mentioned in Refs. \cite{Eschrig2001,Kallin2012,Pustogow2019}, where accidental (near) degeneracy of two pairing states from distinct IRs is required. In our model, the degeneracy comes from system symmetry and survives even under unaxial strain, so the formation of such kind of domains would be more natural and likely.
	
	Despite our results tend to support an odd-to-even OP transition, there is still a possibility that the pairing symmetry keeps unchanged under strain, as indicated by the resent NMR measurements \cite{Pustogow2019}. Comparing Fig. \ref{fig02}(b) with (a), it can be seen that some portions of the phase diagram retain their pairing states when highly strained. In the top left corner the d-wave pairing is always dominant while in the lower right corner the p-wave pairing keeps the most favorable. Note that the $K_{xx}$ reductions of the p- and d-wave OPs are too similar to be distinguished by experiments, a precise measurement of the strain dependence of the out-of-plane Knight shift $K_{zz}$ would help to clarify this issue.

	W.P. Chen thanks J.P. Xiao, F. Xiong, Q.L. Zhu and Y. Zhou for useful discussions.
	This work is supported by NSFC Project No.11874202 and 973 Project No.2015CB921202.

	\section*{Supplemental Materials }
	\begin{table*}
		\caption{The 6 independent BGFs and their relative amplitudes $g_{\Gamma,a}$ of a representative odd(even) pairing state of the unstrained Sr$_2$RuO$_4$, with the SOC strength $\eta=0.2t$. The bold gives amplitudes of the leading components. }
		\begin{tabular}{p{1cm}<{\centering}|p{1cm}<{\centering}|p{2.5cm}<{\centering}|p{2.5cm}<{\centering}|p{1.5cm}<{\centering}|p{1.5cm}<{\centering}|p{2.5cm}<{\centering}|p{2.5cm}<{\centering}}
			\hline
			\hline
			Par&IR&$\bm{d}^{(xz,xz)}+\bm{d}^{(yz,yz)}$&$\bm{d}^{(xz,xz)}-\bm{d}^{(yz,yz)}$&$\bm{d}^{(xy,xy)}$&
			$\bm{d}^{(xz,yz)}$&$d_z^{(xz,xy)}+d_z^{(yz,xy)}$&$\psi^{(xz,xy)}-\psi^{(yz,xy)}$\\
			\hline
			Odd&$B_1$&-0.011&-0.054&-0.064&-0.044& -0.093& $\bm{0.991}$
			\\
			\hline
			\hline
			Par&IR&$\psi^{(xz,xz)}+\psi^{(yz,yz)}$&$\psi^{(xz,xz)}-\psi^{(yz,yz)}$&$\psi^{(xy,xy)}$&$d_{z}^{(xz,yz)}$&$d_{x}^{(xz,xy)}-d_{y}^{(yz,xy)}$&$d_{x}^{(xz,xy)}+d_{y}^{(yz,xy)}$\\
			\hline
			Even&$B_1$&0.003&-0.021&$-\bm{0.441}$&-0.030&$\bm{0.877}$&0.184\\
			\hline
			\hline
		\end{tabular}\label{tab1}
		
	\end{table*}
	
	\begin{table*}
		\caption{The 12 independent BGFs and their relative amplitudes $g_{\Gamma,a}$ of a representative odd(even) pairing state of the strained Sr$_2$RuO$_4$, with the SOC strength $\eta=0.2t$. The bold gives amplitudes of the leading components. }
		\begin{tabular}{p{1cm}<{\centering}|p{1cm}<{\centering}|p{1cm}<{\centering}|p{1cm}<{\centering}|p{1cm}<{\centering}|p{1cm}<{\centering}|p{1cm}<{\centering}|p{1cm}<{\centering}|p{1cm}<{\centering}|p{1cm}<{\centering}|p{1cm}<{\centering}|p{1cm}<{\centering}|p{1cm}<{\centering}|p{1cm}<{\centering}}
			\hline
			\hline
			Par&IR&$d_x^{(xz,xz)}$&$d_x^{(yz,yz)}$&$d_x^{(xy,xy)}$&$d_y^{(xz,xz)}$&$d_y^{(yz,yz)}$&$d_y^{(xy,xy)}$&$d_x^{(xz,yz)}$&$d_y^{(xz,yz)}$&$d_z^{(xz,xy)}$&$d_z^{(yz,xy)}$&$\psi^{(yz,xy)}$&$\psi^{(xz,xy)}$\\
			\hline
			Odd&$A_1$&-0.017&0.017&0.004&-0.029&0.017&0.007&0.026&-0.020&0.089&0.047&$\bm{0.807}$&-$\bm{0.579}$
			\\
			\hline
			\hline                                                                        			Par&IR&$\psi_-^{(xz,xz)}$&$\psi_-^{(yz,yz)}$&$\psi_-^{(xy,xy)}$&$\psi_+^{(xz,xz)}$&$\psi_+^{(yz,yz)}$&$\psi_+^{(xy,xy)}$&$d_{z-}^{(xz,yz)}$&$d_{z+}^{(xz,yz)}$&$d_{x-}^{(xz,xy)}$&$d_{x+}^{(xz,xy)}$&$d_{y-}^{(yz,xy)}$&$d_{y+}^{(yz,xy)}$\\
			\hline
			Even&$A_1$&0.000&-0.022&0.113&0.003&-0.016&-0.015&0.019&-0.003&0.169&-0.011&$\bm{0.964}$&-0.166\\
			\hline
			\hline
		\end{tabular}\label{tab2}
		
	\end{table*}
	\renewcommand\arraystretch{1}

	\subsection{Tight-binding model and Pairing Symmetry analysis}
	
	The FS sheets of Sr$_2$RuO$_4$ are derived from the $t_{2g}$ orbitals $d_{xz}$, $d_{yz}$ and $d_{xy}$ of Ru, so we describe the 2D tight-binding Hamiltonian in a three-orbital representation as:
	\begin{eqnarray}
	H(\bm{k})=\sum_{\bm{k},s} C_s^\dagger(\bm{k}) H_{0s}(\bm{k})C_s(\bm{k})
	,\end{eqnarray}
	where the spinor $C_s(\bm{k})$ is $(c_{xz,\bm{k}s},c_{yz,\bm{k}s},c_{xy,\bm{k}-s})^T$, with $c_{l,\bm{k}s}$ being electron annihilation operator and $s= +1, -1$ denoting the spin $\uparrow, \downarrow$. The $H_{0s}(\bm{k})$ is given by
	\begin{eqnarray}\label{eqn1}
	H_{0s}(\bm{k})=
	\begin{bmatrix}
	\xi_{xz,\bm{k}}& \lambda_{\bm{k}}-i s \eta& i \eta \\
	\lambda_{\bm{k}}+i s \eta & \xi_{yz,\bm{k}} & -s\eta \\
	-i \eta & -s\eta & \xi_{xy,\bm{k}}
	\end{bmatrix}
	,\end{eqnarray}
	with
	\begin{eqnarray}
	\xi_{xz,\bm{k}}&=&-2t(1-b \epsilon_{xx}) \cos k_x-2t^\perp(1-b \epsilon_{yy}) \cos k_y-\mu \nonumber \\
	\xi_{yz,\bm{k}}&=&-2t^\perp (1-b\epsilon_{xx}) \cos k_x-2t (1-b\epsilon_{yy})\cos k_y-\mu \nonumber \\ \xi_{xy,\bm{k}}&=&-2t'[(1-b\epsilon_{xx}) \cos k_x+(1-b\epsilon_{yy})\cos k_y] \nonumber \\
	&&-4t''\cos k_x\cos k_y-\mu' \nonumber \\
	\lambda_{\bm{k}}&=&-4t'''\sin k_x \sin k_y
	,\end{eqnarray}
	where $\eta$ denotes the SOC strength and $\lambda_{\bm{k}}$ the inter-orbital hopping term. $\epsilon_{xx/yy}$ represent the in-plane strain of Sr$_2$RuO$_4$ and are related by the Poisson's ratio through $\epsilon_{yy}=-v_{xy}\epsilon_{xx}$, where $v_{xy}=0.39$. At zero strain, i.e., $\epsilon_{xx}=0$, parameters are set to fit the high-resolution photoemission
	:$(t,t^\perp,t',t'',t''',\mu,\mu',\eta)=(1,0.1,0.8,0.3,0.05,1.0,1.0,0.2)$. $b=9.1$ is chosen to make the first Lifshitz transition happen when $\epsilon_{xx}=\epsilon_{V}\approx-0.75 \%$. All the three energy bands $\alpha, \beta$ and $\gamma$ are two-fold degenerate due to the cooperation of space inversion and time-reversal symmetries.
	
	The BdG Hamiltonian of the superconducting Sr$_2$RuO$_4$ is defined as
	\begin{eqnarray}\label{eqn2}
	H_{BdG}(\bm{k})=
	\begin{bmatrix}
	H(\bm{k}) & \Delta(\bm{k}) \\
	\Delta^\dagger(\bm{k}) & -H^t(\bm{-k})
	\end{bmatrix}
	,\end{eqnarray}
	where $\Delta(\bm{k})$ is a $6\times6$ matrix.
	Because Sr$_2$RuO$_4$ is inversion-symmetric, pairings with different parities in momentum space could not mix together. In the basis of $C(\bm{k})=(C_+(\bm{k}),C_-(\bm{k}))^t$, $\Delta(\bm{k})$ can be divided into four subsectors as
	\begin{eqnarray}
	\Delta(\bm{k})=
	\begin{bmatrix}
	\Delta_{++}(\bm{k}) & \Delta_{+-}(\bm{k}) \\
	\Delta_{-+}(\bm{k}) & \Delta_{--}(\bm{k})
	\end{bmatrix}
	.\end{eqnarray}
	If the even-parity $\Delta(\bm{k})$ are block off-diagonal(diagonal), then the odd-parity ones belonging to the same IR $\Gamma$ must be block diagonal(off-diagonal).  	
	To avoid confusion, we classify the gap functions by ``even/odd'' in momentum space rather than by ``singlet/triplet'' in spin space. As we will present, the former is not equivalent to the latter in a multiple-orbital pairing system, which is distinct from the one-band case. Because of the Fermi statistics, the superconducting gap matrix always obeys
	\begin{eqnarray}\label{eqn5}
	\Delta(\bm{k})=-\Delta^t(-\bm{k})
	.\end{eqnarray} 	
	Denote the $2\times 2$ pairing matrix between orbitals $m,n$ as $\Delta^{m,n}(\bm{k})$
	\begin{eqnarray}\label{eqn7}
	\Delta^{m,n}(\bm{k})=
	\begin{cases}
	[\bm{d}^{m,n}(\bm{k})\cdot \bm{\sigma}]i\sigma_y & {triplet }  \\
	\psi^{m,n}(\bm{k})i\sigma_y & {singlet }
	\end{cases}
	,\end{eqnarray}
	we have $\bm{d}^{m,n}(-\bm{k})=-\bm{d}^{n,m}(\bm{k}),\psi^{m,n}(-\bm{k})=\psi^{n,m}(\bm{k})$.
	If the parity is even in momentum space, for instance, $\bm{d}^{m,n}(\bm{k}), \psi^{m,n}(\bm{k})\sim \cos k_x$, then we have the orbital parity
	\begin{eqnarray}
	\begin{cases}
	\bm{d}^{m,n}(\bm{k})=-\bm{d}^{n,m}(\bm{k}) \\
	\psi^{m,n}(\bm{k})=\psi^{n,m}(\bm{k})
	\end{cases}
	,\end{eqnarray}
	else if the parity is odd, the following orbital parity must be satisfied:
	\begin{eqnarray}
	\begin{cases}
	\bm{d}^{m,n}(\bm{k})=\bm{d}^{n,m}(\bm{k})  \\
	\psi^{m,n}(\bm{k})=-\psi^{n,m}(\bm{k})
	\end{cases}
	.\end{eqnarray}
	
	We present all the nearest-neighbor pairing basis gap functions(BGFs) of the unstrained and strained Sr$_2$RuO$_4$ in Table \ref{tab3} and Table \ref{tab4}, respectively. The BGFs may consist of pairings from different orbital channels and are shown in $\bm{d}$-vector or $\psi$ forms in these tables. In analogy with Eq. (\ref{eqn7}),  the $6\times 6$ matrix $\Delta^{(m,n)}$ can be defined, which gives accordingly $3\times 3$ $\psi^{(m,n)}$ and $\bm{d}$-vector $\bm{d}^{(m,n)}$. The only two nonzero elements of $\psi^{(m,n)}$ are: $[\psi^{(m,n)}]_{mn}=\psi^{m,n}$, $[\psi^{(m,n)}]_{nm}=\psi^{n,m}$, while those of $\bm{d}^{(m,n)}$ can be given similarly. Function such as $\psi[\bm{d}]^{(m,n)}\pm\psi[\bm{d}]^{(m',n')}$ in these tables denotes the two pairings from different orbital channels mixing together symmetrically or antisymmetrically.
	
	The total number of the $\bm{k}$-dependent BGFs of an $L$-orbital system can be empirically expressed as $N^F=n^F\times L^2$, where $n^F$ is the number of the BGFs in the one-band case. Here $n^F=8$ for both $C_{4v}$ and $C_{2v}$, so there are 72 BGFs in each table.
	As the unaxial strain is applied, the point-group symmetry is reduced as $C_{4v} \rightarrow C_{2v}$ and the corresponding IRs are transformed as $\{A_1, B_1\} \rightarrow A_1$, $\{A_2, B_2\} \rightarrow A_2$ and $E\rightarrow \{B_1,B_2\}$ . In the matrix form, each BGF $D_{\Gamma.a}(\bm{k})$ is so normalized that:
	\begin{eqnarray}
	\frac{1}{N}\sum_{\bm{k}} Tr\{D_{\Gamma,a}(\bm{k})D^{\dagger}_{ \Gamma,a}(\bm{k})\}=1
	,\end{eqnarray}
	where $a$ labels the BGF in IR $\Gamma$ and N is the number of unit cells in the system.
	
	Table \ref{tab1} and \ref{tab2} show the relative amplitudes $g_{\Gamma,a}$ of each BGF for two representative pairing states without and under strain, respectively, with the normalization relation $\sum_a |g_{\Gamma,a}|^2=1$. The gap matrix can be expressed as a linear superposition of the BGFs in the IR $\Gamma$:
	\begin{eqnarray}\label{eqn10}
	\Delta(\bm{k})=\Delta\sum_{a}  g_{\Gamma,a}D_{\Gamma,a}(\bm{k})
	,\end{eqnarray}
	where $\Delta$ is the temperature dependent gap size. All $g_{\Gamma,a}$ in these tables are real, thus the obtained OPs here are TRI pairings£¬ indicating
	\begin{eqnarray}\label{eqn11}
	\Delta(\bm{k}) u^\dagger_T=u_T\Delta^\dagger(\bm{k})
	,\end{eqnarray}
	with $u_{T}=i \sigma_y$ being the unitary factor relevant to the time-reversal operator. This relation can also be checked directly from the tables.	
	We now make the following transformation of $\Delta(\bm{k})$,
	\begin{eqnarray}\label{eqn12}
	\Delta(\bm{k})&\rightarrow&U^\dagger(\bm{k}) \Delta(\bm{k}) U^*(-\bm{k}) \nonumber \\
	&=&\begin{bmatrix}
	U^\dagger_{+}(\bm{k}) \Delta_{++}(\bm{k}) U_{+}^*(-\bm{k})& U^\dagger_{+}(\bm{k}) \Delta_{+-}(\bm{k}) U_{-}^*(-\bm{k}) \\
	U^\dagger_{-}(\bm{k}) \Delta_{-+}(\bm{k}) U_{+}^*(-\bm{k}) & U^\dagger_{-}(\bm{k})\Delta_{--}(\bm{k}) U_{-}^*(-\bm{k})
	\end{bmatrix} \nonumber \\
	&=&\begin{bmatrix}
	\Delta_{\Uparrow\Uparrow}(\bm{k}) & \Delta_{\Uparrow\Downarrow}(\bm{k}) \\
	\Delta_{\Downarrow\Uparrow}(\bm{k}) & \Delta_{\Downarrow\Downarrow}(\bm{k})
	\end{bmatrix}
	,\end{eqnarray}
	where $U(\bm{k})=\begin{bmatrix}
	U_{+}(\bm{k})& \\
	& U_{-}(\bm{k})
	\end{bmatrix}$, with $U_{s}(\bm{k})$ being the unitary $3\times 3$ matrix diagonalizing the normal-state Hamiltonian $H_{0s}(\bm{k})$. Here $\bar{s}=\Uparrow,\Downarrow$ denote the up, down pseudo-spin. Then the projected band gaps on the FS will be given by $\Delta^{l,l}_{\bar{s}\bar{s}^{'}}(\bm{k})$, where $l=\alpha,\beta,\gamma$ is the band index. The inter-band pairings are neglected due to the large energy separation between these bands. The corresponding pseudo-$\bm{d}$-vector or $\psi$ on the band $l$ could also be defined through
	\begin{eqnarray}\label{eqn13}
	\Delta^{l,l}_{\bar{s}\bar{s}^{'}}(\bm{k})=
	\begin{cases}
	[(\bm{d}^l\cdot \bm{\sigma})i\sigma_y]_{\bar{s}\bar{s}^{'}} & {triplet}\\
	[\psi^li\sigma_y]_{\bar{s}\bar{s}^{'}} & {singlet}
	\end{cases}
	.\end{eqnarray}		
	\renewcommand\arraystretch{1.5}
	\begin{table*}
		\caption{Same as Table \ref{tab1} but the SOC strength is $\eta=0.3t$.  }
		\begin{tabular}{p{1cm}<{\centering}|p{1cm}<{\centering}|p{2.5cm}<{\centering}|p{2.5cm}<{\centering}|p{1.5cm}<{\centering}|p{1.5cm}<{\centering}|p{2.5cm}<{\centering}|p{2.5cm}<{\centering}}
			\hline
			\hline
			Par&IR&$\bm{d}^{(xz,xz)}+\bm{d}^{(yz,yz)}$&$\bm{d}^{(xz,xz)}-\bm{d}^{(yz,yz)}$&$\bm{d}^{(xy,xy)}$&
			$\bm{d}^{(xz,yz)}$&$d_z^{(xz,xy)}+d_z^{(yz,xy)}$&$\psi^{(xz,xy)}-\psi^{(yz,xy)}$\\
			\hline
			Odd&$B_1$&-0.026&-0.013&-0.022&-0.040& -$\bm{0.581}$& $\bm{0.812}$
			\\
			\hline
			\hline
			Par&IR&$\psi^{(xz,xz)}+\psi^{(yz,yz)}$&$\psi^{(xz,xz)}-\psi^{(yz,yz)}$&$\psi^{(xy,xy)}$&$d_{z}^{(xz,yz)}$&$d_{x}^{(xz,xy)}-d_{y}^{(yz,xy)}$&$d_{x}^{(xz,xy)}+d_{y}^{(yz,xy)}$\\
			\hline
			Even&$B_1$&0.00&-0.025&$-0.093$&-0.032&$\bm{0.990}$&0.099\\
			\hline
			\hline
		\end{tabular}\label{tab5}
		
	\end{table*}
	\renewcommand\arraystretch{1}	
	We now prove that if the original pairing state is TRI, then the projected gaps on the FS will maintain this time-reversal symmetry.
	In the basis of $C(\bm{k})$, $u_T$ can be reexpressed as $u_T=\begin{bmatrix}&U_T \\-U_T &\end{bmatrix}$ with $U_T=\begin{bmatrix}1&& \\ &1& \\&&-1\end{bmatrix}$. The time-reversal invariance of the normal-state Hamiltonian can be written as $u_TH^*(-\bm{k})u_T^\dagger=H(\bm{k})$, from which we get
	\begin{eqnarray}\label{eqn15}
	\begin{cases}
	U_TH^*_{0+}(-\bm{k})U_T=H_{0-}(\bm{k}) \\
	U_{-}(\bm{k})=U_TU^*_{+}(-\bm{k})
	\end{cases}
	.\end{eqnarray}
	For an arbitrary TRI pairing $\Delta(\bm{k})$, the relation (\ref{eqn11}) leads to
	\begin{eqnarray}\label{eqn15-1}
	\begin{cases}
	U_T\Delta_{++}(\bm{k})U_T=-\Delta^\dagger_{--}(\bm{k}) \\
	U_T\Delta_{+-}(\bm{k})U_T=\Delta^\dagger_{+-}(\bm{k})
	\end{cases}
	.\end{eqnarray}	
	Combining  Eqs. (\ref{eqn12}), (\ref{eqn15}) and (\ref{eqn15-1}) yields
	\begin{eqnarray}
	\Delta_{\Uparrow\Uparrow}(\bm{k})&=&U^\dagger_{+}(\bm{k}) \Delta_{++}(\bm{k}) U_{+}^*(-\bm{k}) \nonumber\\
	&=&-U^t_{-}(-\bm{k}) \Delta^\dagger_{--}(\bm{k}) U_{-}(\bm{k}) \nonumber \\
	&=&U^t_{-}(-\bm{k}) \Delta^*_{--}(-\bm{k}) U_{-}(\bm{k}) \nonumber \\
	&=&\Delta^*_{\Downarrow\Downarrow}(-\bm{k}) \nonumber\\
	&=&-\Delta^\dagger_{\Downarrow\Downarrow}(\bm{k}) \label{eqn15-2}\\
	\Delta_{\Uparrow\Downarrow}(\bm{k})
	&=&U^\dagger_{+}(\bm{k}) \Delta_{+-}(\bm{k}) U_{-}^*(-\bm{k}) \nonumber\\
	&=&U^t_{-}(-\bm{k}) \Delta^\dagger_{+-}(\bm{k}) U_{+}(\bm{k})\nonumber \\
	&=&-U^t_{-}(-\bm{k}) \Delta^*_{-+}(-\bm{k}) U_{+}(\bm{k}) \nonumber \\
	&=&-\Delta^*_{\Downarrow\Uparrow}(-\bm{k}) \nonumber \\
	&=&\Delta^\dagger_{\Uparrow\Downarrow}(\bm{k})  \label{eqn15-3}
	.\end{eqnarray}	
	In each equation we used twice of the Eq.(\ref{eqn5}), i.e., $\Delta(\bm{k})=-\Delta^t(-\bm{k})$. These two equations guarantee the $\bm{d}$-vector or $\psi$ defined by Eq.(\ref{eqn13}) to be real, i.e., TRI.
	
	\subsection{self-consistent gap equation}
	
	The temperature dependent gap function can be obtained from the mean-field self-consistent gap equation	
	\begin{eqnarray}\label{eqn16}
	\Delta_{\sigma_1 \sigma_2}(\bm{k})&=& \frac{1}{N}\sum\limits _{\bm{k}^\prime \sigma_3\sigma_4} V_{\sigma_1\sigma_2\sigma_3\sigma_4}(\bm{k},\bm{k}^\prime) \langle c_{\bm{k},\sigma_3}c_{-\bm{k},\sigma_4} \rangle , \nonumber \\
	\end{eqnarray}
	where $\sigma$ is the index combining the three orbitals and two spins, 	
	and the interaction terms $V(\bm{k},\bm{k}^\prime)$ could be expanded by all the BGFs as follows:	
	\begin{eqnarray}
	V_{\sigma_1\sigma_2\sigma_3\sigma_4}&&(\bm{k},\bm{k}^\prime) = \nonumber\\
	&&\sum_{\Gamma a} V_{\Gamma,a} [D_{\Gamma,a}(\bm{k})]_{\sigma_1 \sigma_2} [D^*_{\Gamma,a}(\bm{k}')]_{\sigma_3\sigma_4}
	,\end{eqnarray}
	with $V_{\Gamma,a}$ being the pairing interaction for $\Gamma,a$ channel.
	Multiplying by $[D^{*}_{ \Gamma,a}(\bm{k})]_{\sigma_1,\sigma_2}$ the two sides of equation (\ref{eqn16}), then taking trace over $\sigma_{1,2}$ indexes and making a sum over $\bm{k}$, we have	
	\begin{eqnarray}\label{eqn17}
	\tilde{g}_{\Gamma,a}=\frac{1}{N}\sum\limits _{\bm{k}\sigma_3\sigma_4} V_{\Gamma,a}[D^*_{\Gamma,a}(\bm{k})]_{\sigma_3\sigma_4}\langle c_{\bm{k},\sigma_3}c_{-\bm{k},\sigma_4} \rangle
	,\end{eqnarray}
	where $\tilde{g}_{\Gamma,a}=\Delta(T) \times g_{\Gamma,a}$ is the temperature dependent gap amplitudes of each BGF defined through equation (\ref{eqn10}). Then the gap size $\Delta(T)$ and all these $g_{\Gamma,a}$ could be obtained by solving equation (\ref{eqn17}) through iteration.			
	
	\subsection{ SOC strength $\eta=0.3t$}
\begin{figure}
		\vspace{0.cm}
		\includegraphics[width=8cm]{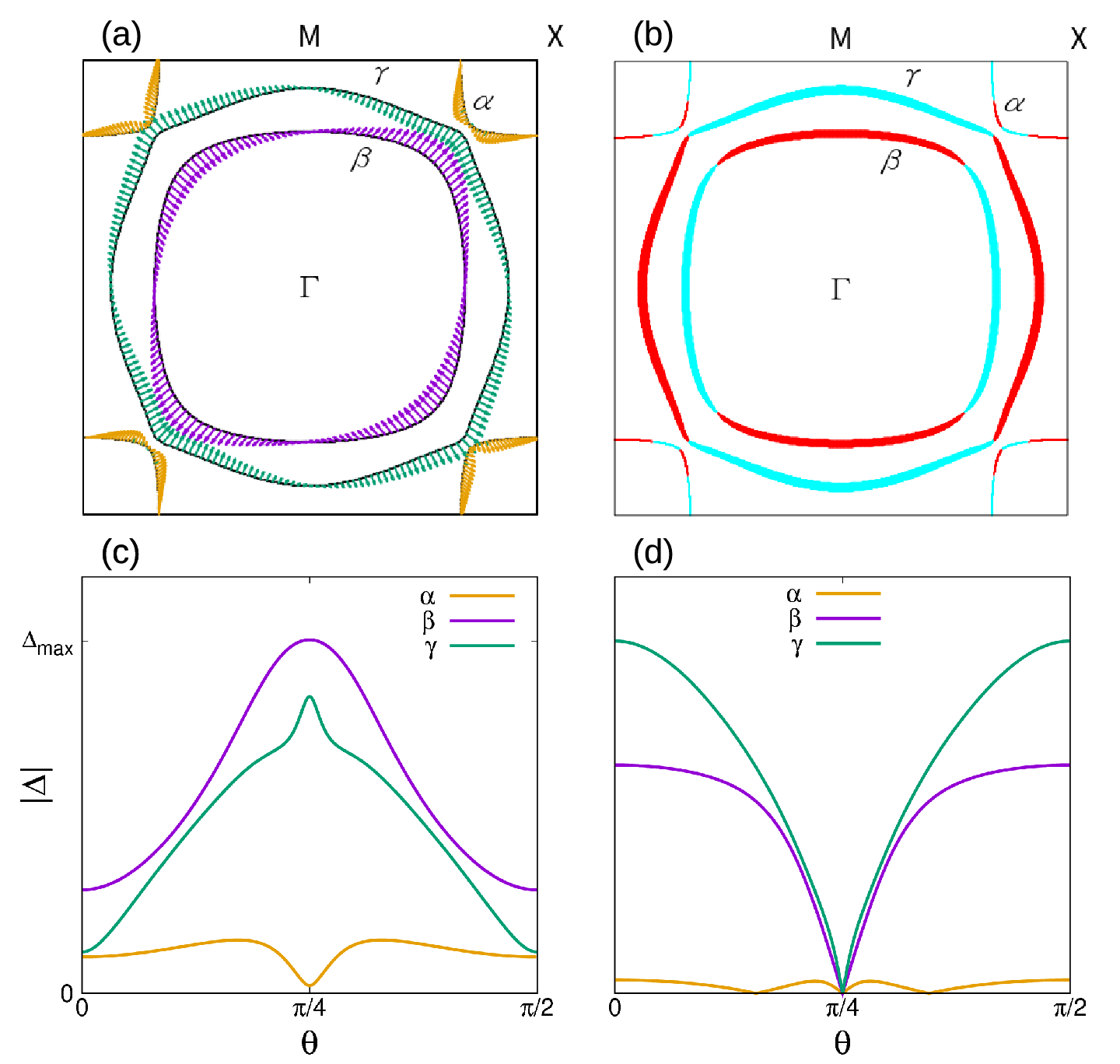}
		\vspace{-0.cm}
		\caption{\label{s01}
			Projected OPs on the FS in the unstrained Sr$_2$RuO$_4$, with $\eta=0.3t$. (a) The projected $\bm{d}$-vectors of the p-wave pairing. (b) The projected $\psi$ of the d-wave pairing, where the red(cyan) line represents the gap sign $-$($+$) and the line width denotes the gap size. Their detailed gap-size dispersions on the FS are displayed in (c)(d).
		}
	\end{figure}

	We also present the p- and d-wave gap structures in the unstrained Sr$_2$RuO$_4$ with a relatively larger SOC strength $\eta=0.3t$ here, as shown in Fig. S\ref{s01}. To keep the FS nearly unchanged, the chemical potential and inter-orbital hopping strength are modified accordingly as $(\mu, \mu', t''')=(1.05, 0.85, 0)t$. The corresponding relative amplitudes $g_{\Gamma,a}$ of each dominant BGF for a representative pairing state are given by Table \ref{tab5}. Compared to the case $\eta=0.2t$, although the ratios between BGFs of the p- or d-wave pairing vary a lot, the leading components are nearly the same, resulting in a similar gap structure. The p-wave pairing still holds minima along the $\langle 100 \rangle$ direction on $\beta$ and $\gamma$, while it's along $\langle 110 \rangle$ on $\alpha$ . It can be seen that this p-wave gap remains nearly nodal with an in-plane projected $\bm{d}$-vector. Gaps of the d-wave pairing are still d-wavelike and dominant on $\beta$ and $\gamma$.
	
	\renewcommand\arraystretch{1.3}
	
	\begin{table*}[t]
		\caption{ The 72 NN BGFs for the 2D point group $C_{4v}$. The 1st column shows the five
			IRs. The symbol $+/-$ in the 2nd and 3rd columns denote the sign given by the basis functions under a fourfold rotation and mirror reflection, respectively. The symbol ``$\pm$" in 2D IR $E$ represents the two degenerate BGFs. The $6\times 6$ BGF can be described by a $3\times 3$ $\bm{d}$-vector or $\psi$, which obeys $\bm{d}^{m,n}(\bm{k})=-\bm{d}^{n,m}(\bm{k}), \psi^{m,n}(\bm{k})=\psi^{n,m}(\bm{k})$ for the even parity and $\bm{d}^{m,n}(\bm{k})=\bm{d}^{n,m}(\bm{k}), \psi^{m,n}(\bm{k})=-\psi^{n,m}(\bm{k})$ for the odd parity. The red(blue) box denotes the dominant BGFs of the even(odd) pairing in Table \ref{tab1}.}
		\begin{tabular}{p{1cm}<{\centering}p{1cm}<{\centering}p{1cm}<{\centering} p{6cm}<{\centering} p{6cm}<{\centering}}
			\hline
			\hline
			\multirow{2}{*}{$C_{4v}$} & \multirow{2}{*}{$C_{4}$} & \multirow{2}{*}{$M_{xz}$} &  \multicolumn{2}{c}{Basis function}\\
			&&&Even& Odd \\
			\specialrule{0.05em}{0pt}{6pt}
			$A_1$&+&+& 	
			
			\tabincell{c}{$\psi^{(xz,xz)}+\psi^{(yz,yz)}\propto \cos k_x +\cos k_y$  \\ $\psi^{(xz,xz)}-\psi^{(yz,yz)}\propto \cos k_x -\cos k_y$ \\ $\psi^{(xy,xy)}\propto \cos k_x +\cos k_y$ \\ $d_{z}^{(xz,yz)}\propto i(\cos k_x +\cos k_y)$ \\ $d_{x}^{(xz,xy)}+d_{y}^{(yz,xy)}\propto i(\cos k_x -\cos k_y)$ \\$d_{x}^{(xz,xy)}-d_{y}^{(yz,xy)}\propto i(\cos k_x +\cos k_y)$}&

			\tabincell{c}{$\bm{d}^{(xz,xz)}+\bm{d}^{(yz,yz)}\propto -\hat{x}\sin k_y+\hat{y}\sin k_x$ \\ $\bm{d}^{(xz,xz)}-\bm{d}^{(yz,yz)}\propto \hat{x}\sin k_y+\hat{y}\sin k_x$ \\$\bm{d}^{(xy,xy)}\propto -\hat{x}\sin k_y+\hat{y}\sin k_x$  \\ $\bm{d}^{(xz,yz)}\propto -\hat{x}\sin k_x+\hat{y}\sin k_y$  \\ $d_z^{(xz,xy)}[\propto \sin k_x]-d_z^{(yz,xy)}[\propto \sin k_y]$  \\ $\psi^{(xz,xy)}[\propto i\sin k_y]+\psi^{(yz,xy)}[\propto i\sin k_x]$} \\
			\specialrule{0.05em}{6pt}{6pt}
			
			$A_2$&+&$-$& 				
			\tabincell{c}{\\ \\ \\ $\psi^{(xz,yz)}\propto \cos k_x -\cos k_y$ \\ $d_{y}^{(xz,xy)}-d_{x}^{(yz,xy)}\propto i(\cos k_x -\cos k_y)$ \\$d_{y}^{(xz,xy)}+d_{x}^{(yz,xy)}\propto i(\cos k_x +\cos k_y)$}&
			
			\tabincell{c}{$\bm{d}^{(xz,xz)}+\bm{d}^{(yz,yz)}\propto \hat{x}\sin k_x+\hat{y}\sin k_y$ \\ $\bm{d}^{(xz,xz)}-\bm{d}^{(yz,yz)}\propto \hat{x}\sin k_x-\hat{y}\sin k_y$ \\$\bm{d}^{(xy,xy)}\propto -\hat{x}\sin k_x-\hat{y}\sin k_y$  \\ $\bm{d}^{(xz,yz)}\propto \hat{x}\sin k_y+ \hat{y}\sin k_x$ \\ $\psi^{(xz,xy)}[\propto i\sin k_x]-\psi^{(yz,xy)}[\propto i\sin k_y]$ \\ $d_z^{(xz,xy)}[\propto -\sin k_y]+d_z^{(yz,xy)}[\propto -\sin k_x]$ } \\
			\specialrule{0.05em}{6pt}{6pt}
			
			$B_1$&$-$&+& 				
			\tabincell{c}{$\psi^{(xz,xz)}+\psi^{(yz,yz)}\propto \cos k_x -\cos k_y$  \\ $\psi^{(xz,xz)}-\psi^{(yz,yz)}\propto \cos k_x +\cos k_y$ \\ $\psi^{(xy,xy)}\propto \cos k_x -\cos k_y$ \\ $d_{z}^{(xz,yz)}\propto i(\cos k_x -\cos k_y)$ \\ $\fcolorbox{red}{white}{$d_{x}^{(xz,xy)}-d_{y}^{(yz,xy)}\propto i(\cos k_x -\cos k_y)$}$ \\$d_{x}^{(xz,xy)}+d_{y}^{(yz,xy)}\propto i(\cos k_x +\cos k_y)$}&

			\tabincell{c}{$\bm{d}^{(xz,xz)}+\bm{d}^{(yz,yz)}\propto \hat{x}\sin k_y+\hat{y}\sin k_x$ \\ $\bm{d}^{(xz,xz)}-\bm{d}^{(yz,yz)}\propto -\hat{x}\sin k_y+\hat{y}\sin k_x$ \\$\bm{d}^{(xy,xy)}\propto \hat{x}\sin k_y+\hat{y}\sin k_x$  \\ $\bm{d}^{(xz,yz)}\propto \hat{x}\sin k_x+\hat{y}\sin k_y$  \\ $d_z^{(xz,xy)}[\propto \sin k_x]+d_z^{(yz,xy)}[\propto \sin k_y]$  \\ $\fcolorbox{blue}{white}{$\psi^{(xz,xy)}[\propto i\sin k_y]-\psi^{(yz,xy)}[\propto i\sin k_x]$}$} \\
			\specialrule{0.05em}{6pt}{6pt}
			
			$B_2$&$-$&$-$&			
			\tabincell{c}{\\ \\ \\ $\psi^{(xz,yz)}\propto \cos k_x +\cos k_y$ \\ $d_{y}^{(xz,xy)}+d_{x}^{(yz,xy)}\propto i(\cos k_x -\cos k_y)$ \\$d_{y}^{(xz,xy)}-d_{x}^{(yz,xy)}\propto i(\cos k_x +\cos k_y)$}&   	
			
			\tabincell{c}{$\bm{d}^{(xz,xz)}+\bm{d}^{(yz,yz)}\propto -\hat{x}\sin k_x+\hat{y}\sin k_y$ \\ $\bm{d}^{(xz,xz)}-\bm{d}^{(yz,yz)}\propto -\hat{x}\sin k_x-\hat{y}\sin k_y$ \\$\bm{d}^{(xy,xy)}\propto \hat{x}\sin k_x-\hat{y}\sin k_y$  \\ $\bm{d}^{(xz,yz)}\propto -\hat{x}\sin k_y+\hat{y}\sin k_x$   \\ $\psi^{(xz,xy)}[\propto -i\sin k_x]+\psi^{(yz,xy)}[\propto -i\sin k_y]$ \\ $\fcolorbox{blue}{white}{$d_z^{(xz,xy)}[\propto \sin k_y]-d_z^{(yz,xy)}[\propto \sin k_x]$}$} \\
			\specialrule{0.05em}{6pt}{3pt}
			
			$E$&& &
			
			\tabincell{c}{\\ \\ \\ $\bm{d}^{(xz,yz)}\propto \pm \hat{x}\cos{k_y}-i \hat{y}\cos{k_x}$  \\ $\bm{d}^{(xz,yz)}\propto \pm \hat{x}\cos{k_x}-i \hat{y}\cos{k_y}$ \\ $\psi^{(xz,xy)}[\propto \cos{k_x}]\pm  \psi^{(yz,xy)}[\propto i\cos{k_y}]$\\ $\psi^{(xz,xy)}[\propto \cos{k_y}]\pm  \psi^{(yz,xy)}[\propto i\cos{k_x}]$ \\ $d_z^{(xz,xy)}[\propto \cos{k_x}]\pm  d_z^{(yz,xy)}[\propto i\cos{k_y}]$  \\ $d_z^{(xz,xy)}[\propto \cos{k_y}]\pm  d_z^{(yz,xy)}[\propto i\cos{k_x}]$}&
			
			\tabincell{c}{$d_z^{(xz,xz)}+ d_z^{(yz,yz)}\propto \sin{k_x}\pm i \sin{k_y}$ \\ $d_z^{(xz,xz)}- d_z^{(yz,yz)}\propto \sin{k_x}\pm i\sin{k_y}$ \\$d_z^{(xy,xy)}\propto \sin{k_x}\pm i\sin{k_y}$\\$d_{z}^{(xz,yz)}\propto \sin{k_x}\pm i\sin{k_y}$  \\ $\psi^{(xz,yz)}\propto \sin{k_x}\pm i\sin{k_y}$  \\ $ d_y^{(xz,xy)}[\propto i\sin k_x]\pm d_x^{(yz,xy)}[\propto \sin k_y]$  \\ $ d_y^{(xz,xy)}[\propto i\sin k_y]\pm d_x^{(yz,xy)}[\propto \sin k_x]$\\ $d_y^{(yz,xy)}[\propto i\sin k_y] \pm d_x^{(xz,xy)}[\propto \sin k_x]$  \\ $ d_y^{(yz,xy)}[\propto i\sin k_x] \pm d_x^{(xz,xy)}[\propto \sin k_y]$} \\
			\specialrule{0.05em}{6pt}{3pt}
			\hline
		\end{tabular}\label{tab3}
		
	\end{table*}
	
	\begin{table*}[t]
		\caption{ The 72 NN BGFs for the 2D point group $C_{2v}$. The 1st column shows the four IRs. The symbol $+/-$ in the 2nd and 3rd columns denote the sign given by the basis functions under a twofold rotation and mirror reflection, respectively. The orbitals involved in the pairing are displayed in the 4th column. The symbol ``$\pm$" in the 5th column represents two different BGFs(s- and d-wave) in the same IR. The $6\times 6$ BGF can be described by a $3\times 3$ $\bm{d}$-vector or $\psi$, which obeys $\bm{d}^{m,n}(\bm{k})=-\bm{d}^{n,m}(\bm{k}), \psi^{m,n}(\bm{k})=\psi^{n,m}(\bm{k})$ for the even parity and $\bm{d}^{m,n}(\bm{k})=\bm{d}^{n,m}(\bm{k}), \psi^{m,n}(\bm{k})=-\psi^{n,m}(\bm{k})$ for the odd parity. The red(blue) box denotes the dominant BGFs of the even(odd) pairing in Table \ref{tab2}.}
		\begin{tabular}{p{1cm}<{\centering}p{1cm}<{\centering}p{1cm}<{\centering}p{2cm}<{\centering} p{4cm}<{\centering} p{6cm}<{\centering}}
			\hline
			\hline
			\multirow{2}{*}{$C_{2v}$} & \multirow{2}{*}{$C_{2}$} & \multirow{2}{*}{$M_{xz}$} & \multirow{2}{*}{Orbital} & \multicolumn{2}{c}{Basis function}\\
			&&&&Even& Odd \\
			\specialrule{0.05em}{0pt}{6pt}
			$A_1$&+&+&
			\tabincell{c}{($d_{xz},d_{xz}$) \\ ($d_{yz},d_{yz}$) \\ ($d_{xy},d_{xy}$)\\ ($d_{xz},d_{yz}$) \\ ($d_{xz},d_{xy}$) \\($d_{yz},d_{xy}$)} &	
			
			\tabincell{c}{$\psi_\pm^{(xz,xz)}\propto \cos k_x \pm \cos k_y$  \\ $\psi_\pm^{(yz,yz)}\propto \cos k_x \pm \cos k_y$ \\ $\psi_\pm^{(xy,xy)}\propto \cos k_x \pm \cos k_y$ \\ $d_{z\pm}^{(xz,yz)}\propto i(\cos k_x \pm \cos k_y)$ \\ $d_{x\pm}^{(xz,xy)}\propto i(\cos k_x \pm \cos k_y)$ \\$\fcolorbox{red}{white}{$d_{y\pm}^{(yz,xy)}\propto i(\cos k_x \pm \cos k_y)$}$}&   		
			\tabincell{c}{$d_x^{(xz,xz)}\propto \sin k_y, d_y^{(xz,xz)}\propto \sin k_x$  \\ $d_x^{(yz,yz)}\propto \sin k_y, d_y^{(yz,yz)}\propto \sin k_x$  \\$d_x^{(xy,xy)}\propto \sin k_y, d_y^{(xy,xy)}\propto \sin k_x$  \\ $d_y^{(xz,yz)}\propto \sin k_y, d_x^{(xz,yz)}\propto \sin k_x$  \\ $\fcolorbox{blue}{white}{$\psi^{(xz,xy)}\propto i\sin k_y$}$, $d_z^{(xz,xy)}\propto \sin k_x$  \\ $d_z^{(yz,xy)}\propto \sin k_y$, $\fcolorbox{blue}{white}{$\psi^{(yz,xy)}\propto i\sin k_x$}$}\\
			\specialrule{0.05em}{6pt}{6pt}
			$A_2$&+&$-$&
			\tabincell{c}{($d_{xz},d_{xz}$) \\ ($d_{yz},d_{yz}$) \\ ($d_{xy},d_{xy}$)\\ ($d_{xz},d_{yz}$) \\ ($d_{xz},d_{xy}$) \\($d_{yz},d_{xy}$)} &	
			
			\tabincell{c}{\\ \\ \\ $\psi_\pm^{(xz,yz)}\propto \cos k_x \pm \cos k_y$ \\ $d_{y\pm}^{(xz,xy)}\propto i(\cos k_x \pm \cos k_y)$ \\$d_{x\pm}^{(yz,xy)}\propto i(\cos k_x \pm \cos k_y)$}&   		
			\tabincell{c}{$d_x^{(xz,xz)}\propto \sin k_x, d_y^{(xz,xz)}\propto \sin k_y$  \\ $d_x^{(yz,yz)}\propto \sin k_x, d_y^{(yz,yz)}\propto \sin k_y$  \\$d_x^{(xy,xy)}\propto \sin k_x, d_y^{(xy,xy)}\propto \sin k_y$  \\ $d_y^{(xz,yz)}\propto \sin k_x, d_x^{(xz,yz)}\propto \sin k_y$  \\ $\psi^{(xz,xy)}\propto i\sin k_x$, $\fcolorbox{blue}{white}{$d_z^{(xz,xy)}\propto \sin k_y$}$  \\ $\fcolorbox{blue}{white}{$d_z^{(yz,xy)}\propto \sin k_x$}$, $\psi^{(yz,xy)}\propto i\sin k_y$} \\
			\specialrule{0.05em}{6pt}{6pt}
			
			$B_1$&$-$&$+$&
			\tabincell{c}{($d_{xz},d_{xz}$) \\ ($d_{yz},d_{yz}$) \\ ($d_{xy},d_{xy}$)\\ ($d_{xz},d_{yz}$) \\ ($d_{xz},d_{xy}$) \\($d_{yz},d_{xy}$)} &	
			
			\tabincell{c}{\\ \\ \\ $d_{x\pm}^{(xz,yz)}\propto i(\cos k_x \pm \cos k_y)$ \\ $d_{z\pm}^{(xz,xy)}\propto i(\cos k_x \pm \cos k_y)$ \\$\psi_{\pm}^{(yz,xy)}\propto \cos k_x \pm \cos k_y$}&   		
			\tabincell{c}{$d_z^{(xz,xz)}\propto \sin k_y$  \\ $d_z^{(yz,yz)}\propto \sin k_y$  \\$d_z^{(xy,xy)}\propto \sin k_y$  \\ $d_z^{(xz,yz)}\propto \sin k_x,\psi^{(xz,yz)}\propto i\sin k_y$  \\ $d_x^{(xz,xy)}\propto \sin k_x, d_y^{(xz,xy)}\propto \sin k_y$  \\ $d_x^{(yz,xy)}\propto \sin k_y, d_y^{(yz,xy)}\propto \sin k_x$} \\
			\specialrule{0.05em}{6pt}{3pt}
			
			$B_2$&$-$&$-$&
			\tabincell{c}{($d_{xz},d_{xz}$) \\ ($d_{yz},d_{yz}$) \\ ($d_{xy},d_{xy}$)\\ ($d_{xz},d_{yz}$) \\ ($d_{xz},d_{xy}$) \\($d_{yz},d_{xy}$)} &	
			
			\tabincell{c}{\\ \\ \\ $d_{y\pm}^{(xz,yz)}\propto i(\cos k_x \pm \cos k_y)$ \\ $\psi_{\pm}^{(xz,xy)}\propto \cos k_x \pm \cos k_y$ \\$d_{z\pm}^{(yz,xy)}\propto i(\cos k_x \pm \cos k_y)$}&   		
			\tabincell{c}{$d_z^{(xz,xz)}\propto \sin k_x$  \\ $d_z^{(yz,yz)}\propto \sin k_x$  \\$d_z^{(xy,xy)}\propto \sin k_x$  \\ $d_z^{(xz,yz)}\propto \sin k_y,\psi^{(xz,yz)}\propto i\sin k_x$  \\ $d_x^{(xz,xy)}\propto \sin k_y, d_y^{(xz,xy)}\propto \sin k_x$  \\ $d_x^{(yz,xy)}\propto \sin k_x, d_y^{(yz,xy)}\propto \sin k_y$} \\
			\specialrule{0.05em}{6pt}{6pt}			
			\hline
		\end{tabular}\label{tab4}		
	\end{table*}

	\renewcommand\arraystretch{1}
	
	\subsection{ degeneracy of the order parameters}
	
	As we know, mixing of distinct BGFs could only occur within a definite IR $\Gamma$.
	However, our calculations reveal that the odd pairings belonging to $B_1$ and $B_2$ for $C_{4v}$ are degenerate and thus could mix together. To understand it, one can define a pseudo-spin-rotation operator as 
	\renewcommand\arraystretch{0.5}
	\begin{eqnarray}
	S_R&=&\begin{bmatrix}
	e^{-i\theta \sigma_z/2}&& \\ &e^{-i \theta \sigma_z/2}& \\ &&e^{-i \theta \bar{\sigma}_z/2}
	\end{bmatrix} \nonumber \\
	&=&\begin{bmatrix}
	e^{-i\theta \sigma_z/2}&& \\ &e^{-i \theta \sigma_z/2}& \\ &&e^{i \theta \sigma_z/2}
	\end{bmatrix} 
	\end{eqnarray}
	\renewcommand\arraystretch{1}in the basis of three orbitals $(d_{xz},d_{yz},d_{xy})$, where the spin of orbital $d_{xy}$ is reversed. Since the pseudo-spin $\overline{S_{z}}$ is conserved, the normal-state Hamiltonian $H(\bm{k})$ is invariant under such a rotation about z axis. In the following, we take $\theta=\pi/2$ to demonstrate the degeneracy.
	
	For the pairings $\Delta^{(m,n)}(\bm{k})$ with $m,n=xz$, $yz$, or $m=n=xy$, by making such a rotation, the gap function $\psi$ and $\bm{d}$-vector will transform as
	\begin{eqnarray}
	\begin{cases}
	d_x \rightarrow \pm d_y,d_y \rightarrow \mp d_x\\	
	d_z,\psi \ \ invariant
	\end{cases}	
	,\end{eqnarray}
	indicating a z-axis $\pi/2$ rotation of the in-plane $\bm{d}$-vector. For instance, the BGF $\bm{d}^{(xy,xy)}\propto\hat{x}\sin k_y+\hat{y}\sin k_x$ of IR $B_1$ for $C_{4v}$ would be transformed to $\bm{d}^{(xy,xy)}\propto-\hat{x}\sin k_x+\hat{y}\sin k_y$ of IR $B_2$. However, the `anti-parallel-spin' pairings which is singlet, or triplet with out-of-plane $\bm{d}$-vector, would be invariant under the rotation.
	
	On the other hand, when $m=xz,yz$ and $n=xy$ or $m=xy$ and $n=xz,yz$, the same rotation will lead to
	\begin{eqnarray}
	\begin{cases}
	\psi \leftrightarrow  d_z \\
	d_x,d_y \ \ invariant
	\end{cases}	
	.\end{eqnarray}
	The parallel-spin pairings keep invariant under this rotation in this case. For the `anti-parallel-spin' pairings, this transformation is highly nontrivial since it is between a singlet pairing $\psi$ and triplet pairing $\bm{d}=\hat{z}d_z$. This realization of transformation for the total spin $S$ of Cooper pairs from $0$ to $1$ is due to the fact that not spin $S_{z}$ but pseudo-spin $\overline{S}_{z}$ is conserved. In this sense, the two components of $B_1$ p-wave pairing $\psi^{(xz,xy)}(i\sin k_y)$, $\psi^{(yz,xy)}(i\sin k_x)$ are degenerate with those of $B_2$ $d_z^{(xz,xy)}(\sin k_y)$, $d_z^{(yz,xy)}(\sin k_x)$, respectively. For the same reason, the corresponding odd pairings of $A_1$ and $A_2$ for $C_{4v}$ are degenerate too. Obviously, one can generalize this conclusion to the strained case, where the odd(even) pairings of $A_1(B_1)$ are degenerate with that of $A_2(B_2)$ for $C_{2v}$.	
	
	Despite the degeneracy between the odd pairings of different IRs discussed above, all the even pairings belonging to different 1D IRs for C$_{4v}$ should be non-degenerate since all these even pairing states are invariant under the rotation. Our detailed calculation of the linearized gap equation has also confirmed this.
	
	\bibliography{ref}	

\end{document}